# The co-evolution of brand effect and competitiveness in evolving networks[*]

Jin-Li Guo (郭进利)

Business School, University of Shanghai for Science and Technology,
Shanghai 200093

**Abstract:** The principle that 'the brand effect is attractive' underlies preferential attachment. Here we show that the brand effect is just one dimension of attractiveness. Another dimension is competitiveness. We firstly develop a general framework that allows us to investigate the competitive aspect of real networks, instead of simply preferring popular nodes. Our model accurately describes the evolution of social and technological networks. The phenomenon which more competitive nodes become richer links can help us to understand the evolution of many competitive systems in nature and society. In general, the paper provides an explicit analytical expression of degree distributions of the network. In particular, the model yields a nontrivial time evolution of nodes' properties and scale-free behavior with exponents depending on the microscopic parameters characterizing the competition rules. Secondly, through theoretical analysis and numerical simulations, it reveals that our model has not only the universality for the homogeneous weighted network, but also the character for the heterogeneous weighted network. Thirdly, the paper also develops a model based on a profit-driven mechanism. It can better describe the observed phenomenon in enterprise cooperation networks. We show that standard preferential attachment, the growing random graph, the initial attractiveness model, the fitness model and weighted networks, can all be seen as degenerate cases of our model.

**Key words:** complex network; weighted network; scale-free network; competitive network; universality
**PACS:** 02.50.–r; 89.75.–k; 89.75.Hc; 89.65.–s

## 1. Introduction

In the past decade, sociologists, physicists, and computer scientists have empirically studied networks in such diverse areas as the World Wide Web (WWW), email networks, social networks, citation networks of academic publications, router networks, etc. It is remarkable that Watts et al. proposed WS model and Barabási et al. proposed an evolving model named BA model [1]. Their research began a new era in the study of complex networks. So far, complex networks have been the subject of an increasingly large interest in the scientific community [1-14]. However, numerous examples convincingly indicate that in real systems a node's connectivity and growth rate does not depend on its age alone. Therefore, the BA model has been improved by many scientists so as to describe real networks better. Bianconi and Barabási have proposed a fitness model in view of the competition phenomenon in evolving process [6]. In addition, Dorogovtsev et al. investigated an initial attractiveness model of directed networks [7]. In the model of Dorogovtsev et al., the attractiveness is a constant. However, different nodes in real networks usually have different attractiveness. For instance, some webpages on the WWW may attract considerably more links than others.

[*] Project supported by the National Natural Science Foundation of China (Grant Nos. 70871082) and the Shanghai First-class Academic Discipline Project (Grant No. S1201YLXK).

The principle that 'the brand effect is attractive' underlies preferential attachment [1,12]. The simplest proxy to the brand effect is a node's connectivity. If new connections are made preferentially to more popular nodes, then the degree distribution of nodes follows power laws [12]. However, the brand effect is just one ingredient of attractiveness; another ingredient is the competitiveness, both the initial attractiveness model and the BBV(Barrat-Barthelemy-Vespignani)model[15,16] are more difficult to describe the mechanism of competition networks. Some documents on the WWW through a combination of good content and marketing acquire a large number of links in a very short time, easily overtaking older websites. In a networked community where users establish links to others, indicating their "trust" for the link receiver's opinion, over time, reviewers in the online review community may receive new incoming trust links and also contribute new reviews, both of which increase their competitiveness to other members of the community [4]. This leads to the formation of a competitive network, with high in-degree individuals being the opinion leaders. If two countries have diplomatic relations, an edge between them can be established, which forms a national network of relationships. However, this network cannot well reflect the competition of the economic and military strength of the two countries. The comprehensive national strength reflects a country's competitiveness. In the evolution of the national network, degrees of nodes are considered together with the comprehensive national strength, based on which a competitive network is formed. How do describe the evolution mechanism of this type of competitive networks?

In the paper, we try to combine the fitness of nodes, with their competitiveness, such as the social skills of an individual, the content of a web page, or the content of a scientific article. We develop one simple model that allows us to investigate the competitive effect of the evolution of a real network. It is found that the degree distribution of the competitive model is universal for weighted networks through theoretical analysis and numerical simulations. Although there have been a large number of models in complex networks, the problem concerning which kind of networks has a broader universality has not been discussed in the study of complex networks. This is an important problem. This study will show that standard preferential attachment [1], the growing random graph[1], the initial attractiveness model, the fitness model and weighted networks, can all be seen as degenerate cases of our model.

The rest of this paper is organized as follows. A model with the co-evolution of the brand effect and the competitiveness is firstly proposed to estimate this competitive aspect of real networks in quantitative terms. Assuming that the existence of the competitiveness modifies the preferential attachment to compete for links, we find that the time dependence of a node's connectivity depends on the competitiveness of the node. An analytical expression for the connectivity distribution of the model is derived by motivating an integral equation. The theoretical results are verified by computer simulation. In section 3, it is analytically deduced that the homogeneous coupling weighted network is not only a special case of the competitive network, but the heterogeneous coupling weighted network is also a special case of that. The theoretical results are verified through numerical simulations. In section 4, we develop a model based on a profit-driven mechanism. It describes that the income of an enterprise cooperation network follows Pareto's distribution, which the distribution is claimed to appear in several different aspects relevant to entrepreneurs and business managers. Finally, a conclusion is in Section 5.

## 2. Competitive networks with the brand effect and the competitiveness

In the last few years, a property frequently identified in complex networks is the "scale-free" property [4]. A network is said to be "scale-free" if its degree distribution follows a power law at least asymptotically [1]. The most widely accepted network growth phenomenon that produces a scale-free network is the preferential attachment process [1]. Interestingly, Yuan et al. found that the Sina microblog network is also a scale-free network [5]. Lu et al. find that whereas network structure-based factors such as preferential attachment and reciprocity are significant drivers of network growth, intrinsic node characteristics such as the number of reviews written and textual characteristics such as objectivity, readability, and comprehensiveness of reviews are also significant drivers of network growth [4].

Some individuals in social networks acquire more social links than others, or on the WWW some webpages attract considerably more links than others. The rate at which nodes in a network increase their connectivity depends on their competitiveness to compete for links. Although our model considers a network that grows through the addition of new nodes such as the creation of new webpages, or the emergence of new companies, it is that a node of the network has its own competitiveness. The evolution of the network is not only to relate to the degree of the node, but also to relate to the competitiveness of that node.

Motivated by the above arguments, the following model combines the brand effect with the competitiveness. To incorporate the competitiveness of nodes，we assign a competitiveness parameter $\xi$ to each node, chosen from a distribution $G(x)$, accounting for the comprehensive strength of companies. In other words, the competitive model is simply constructed as follows.

（ⅰ）*Random growth*: The network starts from initial one with $m_0$ nodes. Suppose that nodes arrive at the system in accordance with a Poisson process having rate $\lambda$. Each node entering the network is tagged its own competitiveness $\xi_i$ and fitness $y_i$, where $\xi_i$ and $y_i$ are taken from given distributions $G(x)$ and $F(y)$, respectively. When a new node is added to the system by time $t$, this new site is connected to $m$ ($m \leq m_0$) previously existing vertices. Where $c = \int x dG(x)$ and $E[y] = \int x dF(x)$ are finite.

(ⅱ) *Preferential attachment*: We assume that the probability that a new node will connect to node $i$ already present in the network depends on the connectivity $k_i(t)$ and on its competitiveness $\xi_i$ of that node, such that

$$\Pi(k_i(t)) = \frac{(by_i + d)k_i(t) + a\xi_i}{\sum_i ((by_i + d)k_i(t) + a\xi_i)}, \quad (1)$$

where $b, d \geq 0$. We call the above model Model$_1$.

If $b = d = 0$, $a = 1$, the mode degenerates the model in Ref.[3]. If $b = d = 0$, $a \neq 0$ and the probability density function of $G(x)$ is $\delta(x-1)$, i.e. all competitiveness are equal to 1, the mode degenerates a growing random graph. Without loss of generality, assuming that $b + d \neq 0$. $t_i$ denotes the time at which the *i*th node is added to the system. $k_i(t)$ denotes the degree of the *i*th node at time $t$. Assuming that $k_i(t)$ is a continuous real variable, the rate at which $k_i(t)$ changes is expected to be proportional to the degree $k_i(t)$. Consequently, $k_i(t)$ satisfies the dynamical equation

$$\frac{\partial k_i(t)}{\partial t} = m\lambda \frac{(by_i + d)k_i(t) + a\xi_i}{\sum_i ((by_i + d)k_i(t) + a\xi_i)} \quad (2)$$

Let

$$B = \lim_{t \to \infty} \frac{b}{\lambda t} \sum_i y_i k_i(t) + 2md + ac \quad (3)$$

Assuming that $N(t)$ represents the total number of nodes that occur by time $t$. By the Poisson process theory, we know $E[N(t)] \approx \lambda t$, thus, for sufficiently large $t$, we have

$$B \approx \frac{1}{\lambda t} \sum_i ((by_i + d)k_i(t) + a\xi_i) \quad (4)$$

Let $A = \frac{B}{m}$. Substituting Eq.(4) into Eq. (2), we can rewrite Equation (2) as

$$\frac{\partial k_i(t)}{\partial t} = \frac{(by_i + d)k_i(t) + a\xi_i}{At} \qquad (5)$$

Since $k_i(t_i) = m$, from Eq.(5), we have

$$k_i(t, y, \xi) = (m + \frac{a\xi_i}{by_i + d})(\frac{t}{t_i})^{\beta(y_i)} - \frac{a\xi_i}{by_i + d}, \qquad (6)$$

where $\beta(y) = \frac{by + d}{A}$

The dynamic exponent $\beta(y)$ is bounded, i.e. $0 < \beta(y) < 1$ because a node always increases the number of links in time ($\beta(y) > 0$) and $k_i(t)$ cannot increases faster than $N(t)$ ($\beta(y) < 1$). Since $2m\lambda t = \sum_i k_i(t) = \int dF(y) \int dG(\xi) \int_0^t k_s(t, y, \xi) \lambda ds$, we obtain that $A$ can be determined by the following integral equation

$$\int (m + \frac{ac}{by + d}) \frac{x}{x - (by + d)} dF(y) = ac \int \frac{1}{by + d} dF(y) + 2m \qquad (7)$$

Eq.(7) is equivalent to

$$m \int \frac{x}{x - (by + d)} dF(y) + ac \int \frac{1}{x - (by + d)} dF(y) - 2m = 0 \qquad (8)$$

Equation (8) is said a characteristic equation of the competitive network.

From Eq.(6), we obtain

$$P\{k_i(t, y, \xi) \geq k\} = P\{t_i \leq (\frac{(by_i + d)m + a\xi_i}{(by_i + d)k + a\xi_i})^{\frac{A}{by_i + d}} t\}, \quad t \gg t_i$$

Notice that the node arrival process is the Poisson process having rate $\lambda$, the time $t_i$ follows a gamma distribution with parameter $(i, \lambda)$, therefore, the probability $P\{k_i(t, y, \xi) < k\}$ can be written as

$$P\{k_i(t, y, \xi) < k\} = e^{-\lambda t (\frac{(by_i + d)m + a\xi_i}{(by_i + d)k + a\xi_i})^{\frac{A}{by_i + d}}} \sum_{l=0}^{i-1} \frac{1}{l!} (\lambda t (\frac{(by_i + d)m + a\xi_i}{(by_i + d)k + a\xi_i})^{\frac{A}{by_i + d}})^l \qquad (9)$$

From Eq. (9), we have

$P\{k_i(t, y, \xi) = k\}$

$$\approx \frac{\lambda t A}{(by_i + d)m + a\xi_i} (\frac{(by_i + d)m + a\xi_i}{(by_i + d)k + a\xi_i})^{\frac{A}{by_i + d} + 1} e^{-\lambda t (\frac{(by_i + d)m + a\xi_i}{(by_i + d)k + a\xi_i})^{\frac{A}{by_i + d}}} \frac{1}{(i-1)!} \left[ \lambda t (\frac{(by_i + d)m + a\xi_i}{(by_i + d)k + a\xi_i})^{\frac{A}{by_i + d}} \right]^{i-1} \qquad (10)$$

From Eq.(10), we obtain the stationary average degree distribution

$$P(k) \approx A \int dF(y) \int \frac{1}{(by+d)m+a\xi} \left( \frac{(by+d)m+a\xi}{(by+d)k+a\xi} \right)^{\frac{A}{by+d}+1} dG(\xi), \quad (11)$$

where $A$ is a solution of the characteristic equation (8).

When $b=0$, $d=1$

$$P(k) \approx \int \frac{2+ac/m}{m+a\xi} \left( \frac{m+a\xi}{k+a\xi} \right)^{3+ac/m} dG(\xi) \quad (12)$$

When $a=d=0$, $b=1$, from Eq.(1), Model$_1$ reduces to the fitness model[6]. Using Eq.(11) and the characteristic equation (8), we obtain

$$P(k) \approx \frac{C}{m} \int \frac{1}{y} \left( \frac{m}{k} \right)^{\frac{C}{y}+1} dF(y) \quad (13)$$

where $C$ is a solution of the following integral equation

$$\int \frac{x}{x-y} dF(y) = 2.$$

We consider the simple version of Model$_1$.

(ⅰ) *Random growth*: The network starts from initial one with $m_0$ nodes. Suppose that nodes arrive at the system in accordance with a Poisson process having rate $\lambda$. Each node entering the network is tagged its own $\eta_i$, and assume that $\eta_i$ are independent random variables taken from a given distribution $F(y)$, When a new node is added to the system at time $t$, this new site is connected to $m$ $(m \leq m_0)$ previously existing vertices. Where $c = \int y dF(y)$ is finite.

(ⅱ) *Preferential attachment*: We assume that the probability that a new node will connect to node $i$ already present in the network depends on the connectivity $k_i(t)$ and on the competitiveness $\eta_i$ of that node, such that

$$\Pi(k_i(t)) = \frac{(b\eta_i+d)k_i(t)+a\eta_i}{\sum_i ((b\eta_i+d)k_i(t)+a\eta_i)}, \quad (14)$$

Where $b,d \geq 0$, and $b+d \neq 0$. We call this model Model$_2$.

Similarly to model$_1$, the following is obtained

$$P\{k_i(t,\eta) < k\} = e^{-\lambda t \left( \frac{(b\eta_i+d)m+a\eta_i}{(b\eta_i+d)k+a\eta_i} \right)^{\frac{A}{b\eta_i+d}}} \sum_{l=0}^{i-1} \frac{1}{l!} (\lambda t (\frac{(b\eta_i+d)m+a\eta_i}{(b\eta_i+d)k+a\eta_i})^{\frac{A}{b\eta_i+d}})^l \quad (15)$$

From Eq. (15), we get

$$P(k) \approx \int \frac{A}{(b\eta+d)m+a\eta} \left( \frac{(b\eta+d)m+a\eta}{(b\eta+d)k+a\eta} \right)^{\frac{A}{b\eta+d}+1} dF(\eta), \quad (16)$$

where $A$ is a solution of the characteristic equation (8).

When $b=0$, $d=1$, and the probability density function of $F(x)$ is $\delta(x-1)$, i.e. all competitiveness are equal, the network reduces to the initial attractiveness model. Therefore, from Eq.(16), the initial attractiveness model is a scale-free network with the degree distribution

$$P(k) \approx \frac{2m+a}{m(m+a)}(\frac{m+a}{k+a})^{3+\frac{a}{m}}, \qquad (17)$$

and the degree exponent

$$\gamma = 3 + \frac{a}{m}. \qquad (18)$$

If $F(x)$ is a uniform distribution on $[0,1]$, the characteristic equation (8) reduces to

$$(mA+ac)\ln\left|\frac{A-d}{A-(b+d)}\right| - 2bm = 0 \qquad (19)$$

Eq. (19) is equivalent to

$$(mA+ac)\ln\left(\frac{A-d}{A-(b+d)}\right)^2 - 4bm = 0 \qquad (20)$$

The simulation result of the degree distribution of Model$_2$ with the uniform distribution is shown in Fig. 1. The simulation result is in good agreement with the analytical one.

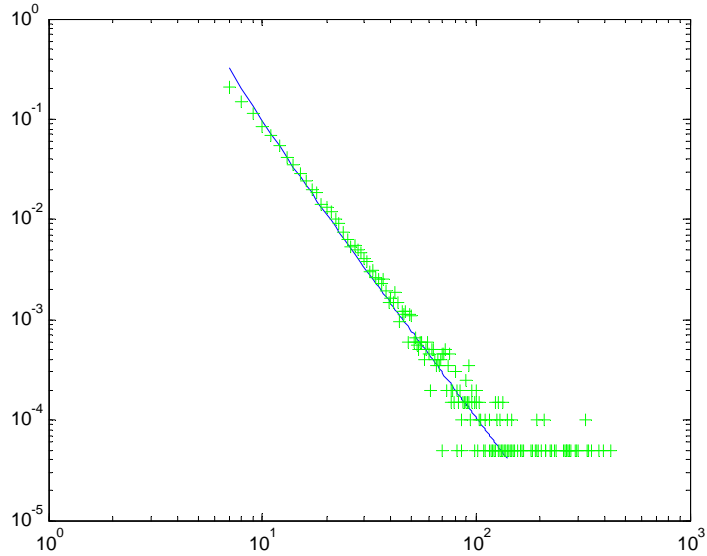

**Fig. 1**. The degree distribution of Model$_2$ with the uniform distribution. The blue line is the theoretical prediction, the symbols represent the simulation result. the number of network nodes is 20000，$m_0 = 12$，$m=7$，$a=5$，$b=2$，$d=1$.

## 3. The universality of the competitive network for the weighted network

In the Internet, it is easy to realize that the introduction of a new connection to a router corresponds to an increase in the traffic handled on the other router's links [16]. Indeed in many technological, large infrastructure and social networks it is commonly believed that a reinforcement of the weights is due to the network's growth. In this spirit, Barrat et al. developed a model for a growing weighted network [16]. They took into account the coupled evolution in time of topology and weights.

The definition of the BBV model is based on two coupled mechanisms: the topological growth and the weights' dynamics. Weighted networks are usually described by an adjacency matrix $w_{ij}$ which represent the weight on the edge connecting nodes $i$ and $j$, with $i,j = 1,2,\ldots,N$, where $N$ is the size of the network. Here undirected networks are only considered, in which the weight matrix is symmetric, that is $w_{ij} = w_{ji}$. The BBV weighted network model is defined as follows [15-16]:

（ⅰ） Random g*rowth*: The network starts from an initial seed of $m_0$ nodes connected by edges with assigned weight $w_0$. Suppose that nodes arrive at the system in accordance with a Poisson process having rate $\lambda$. A new node $n$ is added at time $t$. This new site is connected to $m$ ($m \leq m_0$) previously existing nodes (i.e., each new node will have initially exactly $m$ edges, all with equal weight $w_0$).

（ⅱ） *strength driven attachment*: The new node $n$ preferentially chooses sites with large strength; i.e., node $i$ is chosen according to the probability:

$$\Pi_{n \to i} = \frac{s_i}{\sum_j s_j}, \tag{21}$$

where $s_i = \sum_{j \in \Omega(i)} w_{ij}$ is the strength of node $i$, the sum runs over the set $\Omega(i)$ of the neighbors of $i$.

(ⅲ) *Weights' dynamics*: The weight of each new edge $(n,i)$ is initially set to a given value $w_0$. A new edge on node $i$ will trigger only local rearrangements of weights on the existing neighbors $j \in \Omega(i)$, according to the simple rule

$$w_{ij} \to w_{ij} + \Delta w_{ij}, \tag{22}$$

where $\Delta w_{ij} = \delta_i \dfrac{w_{ij}}{s_i}$ and $\delta_i$ is defined as updating coefficient and it is independent on the time $t$.

This rule yields a total strength increase for node $i$ of $w_0 + \delta_i$, implying that, $s_i \to s_i + w_0 + \delta_i$. After the weights have been updated, the growth process is iterated by introducing a new node, i.e., going back to step ( i ) until the desired size of the network is reached.

The changed strength is composed by three parts: the original strength, the new edge weight brought by the new node and the increment of the old edge weight.

**3.1. Heterogeneous coupling**

In this section, we will focus on the heterogeneous coupling with $\delta_i$ that we will assume are some sample observations which are individually taken from a population $X$ with a distribution $F(x)$, and $E[X] = \int x dF(x) = c$ is finite.

When a new node $n$ is added to the network, an already present node $i$ can be affected in two ways: (i) It is chosen with probability (21) to be connected to $n$; then its connectivity increases by 1, and its strength by $w_0 + \delta_i$. (ii) One of its neighbors $j \in \Omega(i)$ is chosen to be connected to $n$. Then the connectivity of $i$ is not modified but $w_{ij}$ is increased according to the rule Eq. (22), and thus $s_i$ is increased by $\delta_j \dfrac{w_{ij}}{s_j}$. This dynamical process modulated by the respective occurrence probabilities $\dfrac{s_i(t)}{\sum_l s_l(t)}$ and $\dfrac{s_j(t)}{\sum_l s_l(t)}$ is thus described by the following evolution equations for $s_i(t)$ and $k_i(t)$:

$$\frac{ds_i}{dt} = m(w_0 + \delta_i)\frac{s_i}{\sum_j s_j} + \sum_{j \in \Omega(i)} m \frac{s_j}{\sum_l s_l} \delta_j \frac{w_{ij}}{s_j} \tag{23}$$

$$\frac{dk_i}{dt} = m \frac{s_i}{\sum_j s_j}, \tag{24}$$

Since node $j$ is the neighbor of node $i$, and in the renewal process of the network the weight of $i$ and the weight of $j$ have a renewal occur simultaneity, without loss of generality, we can use $\delta_j \approx \delta_i$ in $\sum_{j \in \Omega(i)} m \frac{s_j}{\sum_l s_l} \delta_j \frac{w_{ij}}{s_j}$, therefore

$\sum_{j \in \Omega(i)} m \frac{s_j}{\sum_l s_l} \delta_j \frac{w_{ij}}{s_j} = m \delta_i \frac{s_i}{\sum_l s_l}$. The following is obtained

$$\frac{ds_i}{dt} = m(w_0 + 2\delta_i) \frac{s_i}{\sum_j s_j} \tag{25}$$

Substituting Eq. (24) into Eq. (25) yields:

$$\frac{ds_i}{dt} = (w_0 + 2\delta_i) \frac{dk_i}{dt}$$

Since node $i$ arrives at the network by time $t_i$, we have $k_i(t_i) = m$ and $s_i(t_i) = mw_0$, then the above equation is integrated from $t_i$ to $t$, The following is obtained

$$s_i = (w_0 + 2\delta_i)k_i - 2\delta_i m, \tag{26}$$

and probability (21) is modified as:

$$\Pi_{n \to i} = \frac{(w_0 + 2\delta_i)k_i - 2\delta_i m}{\sum_j [(w_0 + 2\delta_j)k_j - 2\delta_j m]} \tag{27}$$

By comparing probability (27) and probability (14), we know that the weighted network is Model$_2$ with $b = 2, d = w_0, a = -2m$. From Eq.(16), we obtain the stationary average degree distribution of the BBV model

$$P(k) \approx \frac{A}{w_0 m} \int \left( \frac{w_0 m}{(2\eta + w_0)k - 2m\eta} \right)^{\frac{A}{2\eta + w_0} + 1} dF(\eta), \tag{28}$$

where $A$ is a solution of the following integral equation

$$\int (1 - \frac{2c}{2y + w_0}) \frac{A}{A - (2y + w_0)} dF(y) = 2 - 2c \int \frac{1}{2y + w_0} dF(y), \tag{29}$$

According to Eq. (26) and Eq. (15), we obtain that the density function of the strength is:

$$f(x) \approx \frac{A}{w_0 m} \int \frac{1}{2y + w_0} (\frac{w_0 m}{x})^{\frac{A}{2y + w_0} + 1} dF(y) \tag{30}$$

If $F(x)$ is a uniform distribution on $[0,1]$, Eq. (29) reduces to

$$(A - 1) \ln \left| \frac{A - w_0}{A - 2 - w_0} \right| - 4 = 0 \tag{31}$$

Eq.(28) reduces to

$$P(k) \approx \frac{A}{w_0 m} \int_0^1 \left( \frac{w_0 m}{(2\eta + w_0)k - 2m\eta} \right)^{\frac{A}{2\eta + w_0}+1} d\eta \tag{32}$$

Eq.(30) reduces to

$$f(x) \approx \frac{A}{w_0 m} \int_0^1 \frac{1}{2y + w_0} \left( \frac{w_0 m}{x} \right)^{\frac{A}{2y+w_0}+1} dy \tag{33}$$

If $w_0 = 1$, Eq.(31) reduces to

$$(A-1)\ln\left|\frac{A-1}{A-3}\right| - 4 = 0 \tag{34}$$

Let

$$\varphi(x) = (x-1)\ln\left|\frac{x-1}{x-3}\right| - 4. \tag{35}$$

A solution of equation (34) is equivalent to seek an intersection point of $y = \varphi(x)$ and $x$ axis.

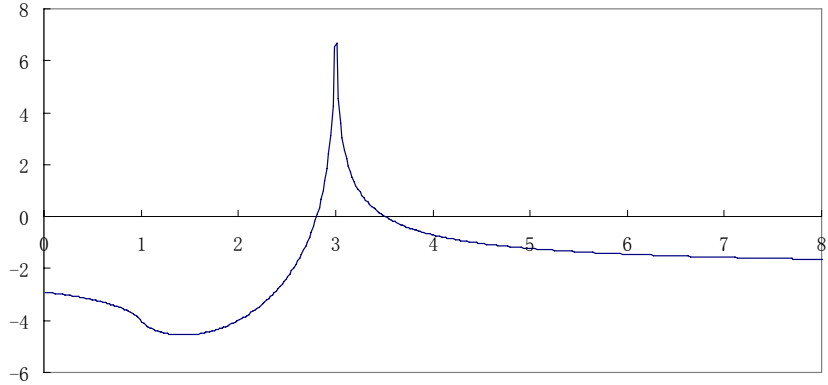

**Fig. 2**. An image of $y = \varphi(x)$.

From Fig. 2, we know that there exist a positive solution of Eq. (34), and then we use numerical integral method to calculate Eq.(32) by solving Eq.(34). The degree distribution of the heterogeneous coupling weighted network is shown in Fig. 3.

The degree distributions both the weighted network and the competitive network corresponding to this weighted network are clearly shown in Fig. 3. The strength distribution of the weighted network is shown in Fig. 4. The validity of the above analysis is verified.

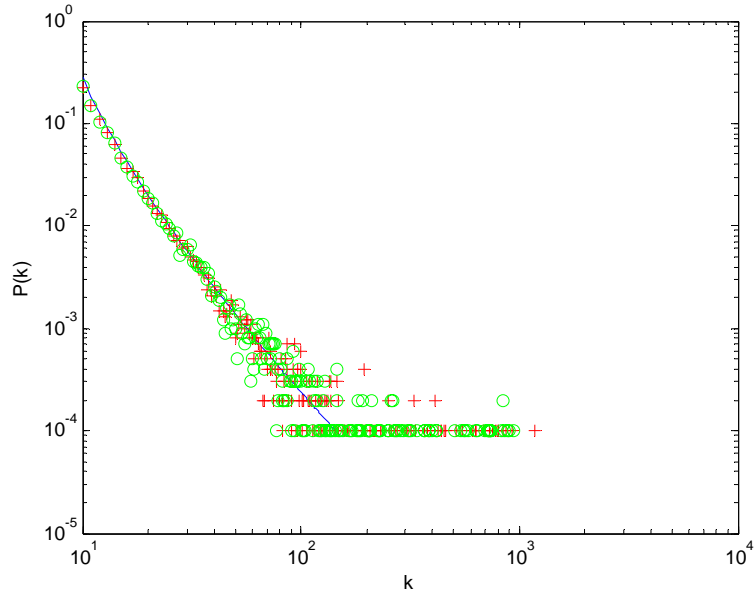

**Fig. 3**. The degree distribution of the heterogeneous coupling weighted network and the competitive network corresponding to this weighted network. The blue line is the theoretical prediction, the plus symbols denotes the simulation of the heterogeneous coupling weighted network with $w_0 = 1$. The circles are the simulation of the competitive network. The number of network nodes are 10000, $m_0 = 12$, $m = 10$ respectively.

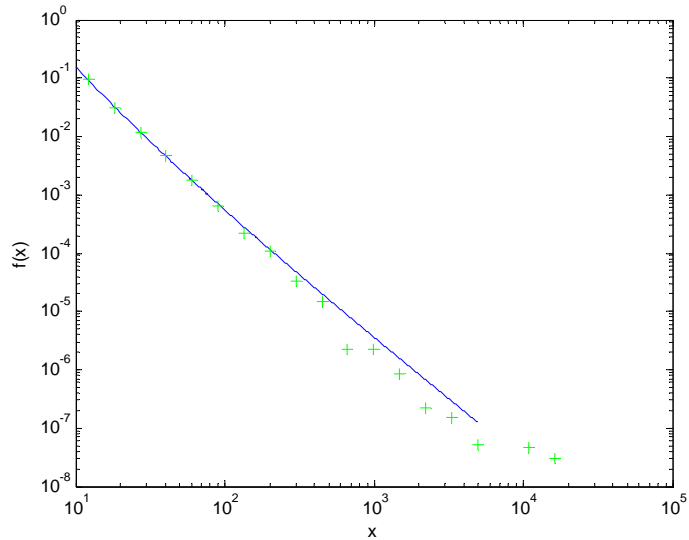

**Fig.4**. The density function of the strength of the heterogeneous coupling weighted network using logarithmic binning method. The blue line is the theoretical prediction; the plus symbols are the simulation. The number of network nodes are 10000, $m_0 = 12$, $m = 10$, $w_0 = 1$ respectively.

### 3.2. Homogeneous coupling

In this section, we will focus on the simplest form of coupling with $\delta_i = \delta$ =const. This case amounts to a very homogeneous system in which all the vertices have an identical coupling between the addition of new edges and the corresponding weights' increase.

Similarly to the heterogeneous coupling, we have

$$s_i = (w_0 + 2\delta)k_i - 2\delta m, \tag{36}$$

and probability (21) is modified as:

$$\Pi_{n \to i} = \frac{k_i - \frac{2\delta}{(w_0 + 2\delta)}m}{\sum_j \left[ k_j - \frac{2\delta}{(w_0 + 2\delta)}m \right]} \tag{37}$$

By comparing probability (37) and probability (14) with $b=0, d=1$, and the probability density function of $F(y)$ is $\delta(y-1)$, it can be inferred that only if

$$a = -\frac{2\delta}{w_0 + 2\delta}m, \tag{38}$$

The probability of the preferential attachment can be modified as $\Pi_{n \to i} = \frac{k_i + a}{\sum_j (k_j + a)}$, which is in accord with that of the initial attractiveness model. That is to say, if the updating coefficient $\delta$ is a constant, the corresponding competitiveness $a$ will also be a constant, which verifies the universality on the competitive network for the weighted network.

Moreover, from Eq.(17), the degree distribution of the BBV weighted network behaves as $P(k) \propto k^{-\gamma}$ where

$$\gamma = 3 + \frac{a}{m} = 2 + \frac{w_0}{2\delta + w_0}. \tag{39}$$

Therefore, the degree distribution of the weighted network can be obtained directly from the results of the competitive network.

According to Eq. (36) and Eq. (15), the following is obtained

$$P\{s_i < x\} = P\{k_i < \frac{x + 2\delta m}{w_0 + 2\delta}\} = e^{-\lambda t (\frac{mw_0}{x})^{2+\frac{a}{m}}} \sum_{l=0}^{i-1} \frac{1}{l!} (\lambda t (\frac{mw_0}{x})^{2+\frac{a}{m}})^l \tag{40}$$

Hence, the density function of $s_i$ is

$$f_{s_i}(x) \approx (2+\frac{a}{m})\frac{(w_0 m)^{2+\frac{a}{m}} \lambda t}{x^{3+\frac{a}{m}}} e^{-\lambda t(\frac{mw_0}{x})^{2+\frac{a}{m}}} \frac{1}{(i-1)!}\left[\lambda t(\frac{mw_0}{x})^{2+\frac{a}{m}}\right]^{i-1} \quad (41)$$

Then the density function of the stationary average node strength distribution can be deduced from Eq. (41):

$$f(x) \approx (2+\frac{a}{m})\frac{(w_0 m)^{2+\frac{a}{m}}}{x^{3+\frac{a}{m}}} \quad (42)$$

By instituting Eq. (38) into the above equation, the density function $f(x)$ can be analogously calculated yielding the power-law behavior

$$f(x) = (\gamma-1)(w_0 m)^{\gamma-1}\frac{1}{x^\gamma}, \quad \gamma = 2 + \frac{w_0}{2\delta + w_0} \quad (43)$$

We take $N=10000$, $m_0=10$, $m=6$ for both networks. The updating coefficient $\delta$ of the weighted network equals to 1, the corresponding the competitiveness $a$ equals to -8/3. The comparisons degree distributions are shown in Fig. 5. The strength distribution of the weighted network is shown in Fig. 6

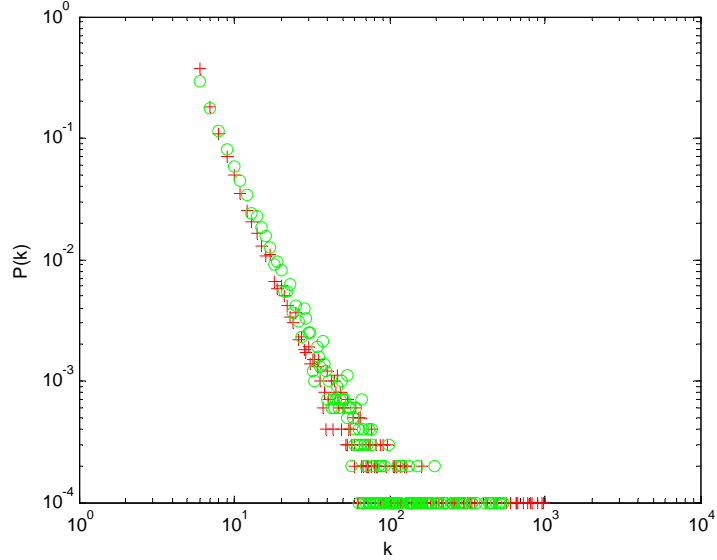

**Fig. 5**. The comparisons of degree distributions of the weighted network and the competitive model. The plus symbols denotes the simulation results of the homogeneous coupling weighted network with $w_0=1$. The circles are the simulation of the competitive network. The number of network nodes are 10000, $m_0=10$, $m=6$, $w_0=1$ respectively.

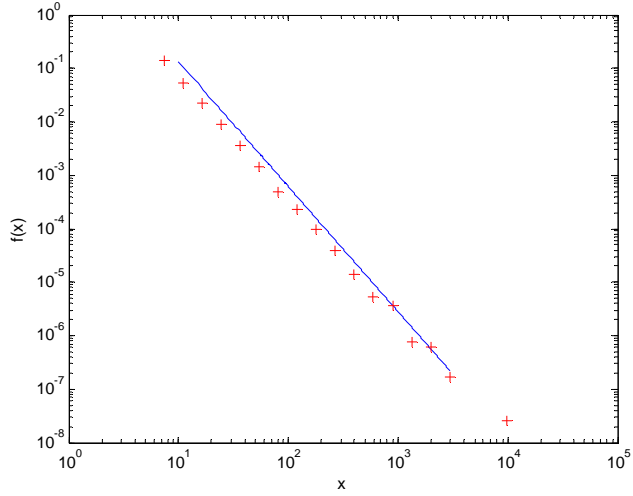

**Fig. 6.** The density function of the strength of the homogeneous coupling weighted network using logarithmic binning method. The blue line is the theoretical prediction; the plus symbols denote the simulation result. The number of network nodes are 10000, $m_0 = 10$, $m = 6$, $w_0 = 1$ respectively.

## 4. Profit-driven evolution of networks

　　Hanaki et al. studied the problem of cooperative behavior emerging in an environment where individual behaviors and interaction structures coevolve [17]. If a company is regarded as a node, an edge between them can be established as two companies have cooperative relationship, which forms an enterprise cooperation network. Each enterprise has the initial investment to join into the network. They can make a profit when companies cooperate. The network is characterized by profit-driven network growth. The definition of a model is based on the profit-driven mechanism: (ⅰ) *Random growth*: The network starts from initial one with $m_0$ nodes. Suppose that nodes arrive at the system in accordance with a Poisson process having rate $\lambda$. Each node entering the system is tagged its own investment $\xi_i$, and suppose that $\xi_i$ are independent random variables having a common distribution $G(x)$, where $c = \int x dG(x)$ is finite. When a new node is added to the system by time $t$, this new site is connected to $m$ ($m \leq m_0$) previously existing vertices. If node $i$ gets a link, it makes profit $\eta_i$, and assume also that $\eta_i$ are independent random variables having a common distribution $F(x)$. (ⅱ) *Preferential attachment*: We assume that the probability that a new node will connect to node $i$ already present in the network depends on total profit $f_i(t)$ of node $i$ and on its investment $\xi_i$ of that node, such that

$$\Pi_{n \to i} = \frac{f_i(t) + \xi_i}{\sum_i (f_i(t) + \xi_i)}, \tag{44}$$

We call the above model Model$_3$.

When a new node $n$ is added to the network, an already present node $i$ can be affected in a way: It is chosen with probability (44) to be connected to $n$; then its connectivity increases by 1, and its profit by $\eta_i$. This dynamical process modulated by the respective occurrence probabilities (44) is thus described by the following evolution equations for $f_i(t)$ and $k_i(t)$:

$$\frac{\partial f_i(t)}{\partial t} = m \eta_i \lambda \frac{f_i(t) + \xi_i}{\sum_i (f_i(t) + \xi_i)} \tag{45}$$

$$\frac{\partial k_i(t)}{\partial t} = m \lambda \frac{f_i(t) + \xi_i}{\sum_i (f_i(t) + \xi_i)} \tag{46}$$

Substituting Eq. (45) into Eq. (46) yields:

$$\frac{df_i}{dt} = \eta_i \frac{dk_i}{dt}$$

Since node $i$ arrives at the network by time $t_i$, we have $k_i(t_i) = m$ and $f_i(t_i) = m\eta_i$, then the above equation is integrated from $t_i$ to $t$, The following is obtained

$$f_i(t) = \eta_i k_i(t), \tag{47}$$

and the probability (44) is modified as:

$$\Pi_{n \to i} = \frac{\eta_i k_i + \xi_i}{\sum_j (\eta_j k_j + \xi_j)}. \tag{48}$$

By comparing probability (48) and probability (1), we know that this network is Model$_1$ with $b = 1, d = 0, a = 1$. From Eq.(11), we obtain the stationary average degree distribution of the model$_3$

$$P(k) \approx A \int dF(y) \int \frac{1}{my + \xi} \left( \frac{my + \xi}{ky + \xi} \right)^{\frac{A}{y}+1} dG(\xi), \tag{49}$$

where $A$ is a solution of the following characteristic equation

$$m\int \frac{x}{x-y} dF(y) + c\int \frac{1}{x-y} dF(y) - 2m = 0 \qquad (50)$$

According to Eq. (47) and Eq. (9), we obtain that the density function of the profits is:

$$g(x) = A\int dF(\eta) \int \frac{1}{\eta(m\eta+\xi)} \left(\frac{m\eta+\xi}{x+\xi}\right)^{\frac{A}{\eta}+1} dG(\xi) \qquad (51)$$

If the probability density function of $F(x)$ and $G(x)$ are $\delta(x-1)$ and $\delta(x-a)$, respectively, then, from Eq.(49) and Eq.(51), the network is a scale-free network with the degree distribution

$$P(k) \approx \frac{2m+a}{m(m+a)} \left(\frac{m+a}{k+a}\right)^{3+\frac{a}{m}},$$

the density function of the profits is

$$g(x) \approx \frac{2m+a}{m(m+a)} \left(\frac{m+a}{x+a}\right)^{3+\frac{a}{m}}, \qquad (52)$$

and the degree exponent $\gamma = 3 + \frac{a}{m}$. Eq.(52) is a generalized Pareto distribution, it shows that a financial wealth of the network follows Pareto's law (also known as the 80–20 rule, the law of the vital few).

## 5. Conclusion

In the recent decade, the preferential attachment based node degrees have considered. However, the co-evolution of networks and behavior has not received as much attention as it deserves. We show that whereas phenomena highlighted in the extant literature, such as preferential attachment and reciprocity, are important drivers of network growth, intrinsic properties of nodes, such as the economic and military strength in the national network of relationships, are also very significant drivers of network growth. Our models admit that the competitiveness and the survival of the fittest are important drivers of network growth. The paper not only provides theoretical proofs, but also simulations. The results from our simulations show that the numerical simulations of the models agree with the analytical results well. The results from our theoretical analysis show that the characteristic equation can serve as a very effective method when dealing with multiscaling network. We also notice the interesting work in Ref.[18]. However, the model and the characteristic equation of Ref. [18] are not all identical with ours.

The competitive networks and the weighted networks are two kinds of evolving models. The evolving mechanism of them is discussed in the paper. The weighted network models can be divided into two categories according to the cases that weight is assigned with fixed or variable values. The former includes the weighted scale-free (WSF) model [13-14] and the Zheng-Trimper-Zheng-Hui (ZTZH) model [19], and the latter includes the BBV model [15-16] and the Dorogovtsev-Mendes(DM) [20], etc. The topological structure of the WSF model or the ZTZH model completely agrees with that of the BA model. Therefore, the two models can be seen as a special case of our model. For the ZTZH expanded model discussed in Ref. [19], its preferential attachment mechanism is in accordance with the fitness model; therefore, it is also a special case of our model. The probability of preferential attachment in the BBV model depends on the node strength. In the paper, it reveals that our model has not only the universality for the homogeneous weighted network, but also the character for the heterogeneous weighted network.

In the economic system, a famous wealth distribution is the phenomenon of Pareto's law. In many countries and regions have found the similar phenomenon since Pareto developed the principle. The cooperation in the economic system is common behavior. The profit-driven model can better describe the observed phenomenon in the economic system.

The paper also shows that standard preferential attachment, the growing random graph, the initial attractiveness model, the fitness model and weighted networks, can all be seen as degenerate cases of our model.

**References**
[1] Albert R and Barabasi A L 2002 *Rev. Mod. Phys.* **74** 47
[2] Beiró M G, Buscha J R and Grynberg S P 2013 *Physica A* **392** 2278
[3] Guo J L, Fan C and Ji Y 2015 *J Syst Sci Complex* (To be published) arXiv:1101.1638
[4] Lu Y D, Jerath K and Singh P V 2013 *Management Science* **59(8)** 1783
[5] Yuan W G, Liu Y, Cheng J J and Xiong F 2013 *Acta Phys. Sin.* **62 (3)** 038901
[6] Bianconi G, Barabasi A L 2001 *Europhys Lett.* **54 (4)** 436
[7] Dorogovtsev S N, Mendes J F F and Samukhin A N 2000 *Phys. Rev. Lett.* **85** 4633
[8] Hu H B, Guo J L and Chen J 2012 Chin. Phys. B 21(11) 118902
[9] Wang J Y, Zhang H G, Wang Z S and Liang H J 2013 Chin. Phys. B 22(9) 090504
[10] Gao Z K, Hu L D and Jin N D 2013 Chin. Phys. B 22(5) 050507
[11] Ling X, Hu M B, Long J C, Ding J X and Shi Q 2013 Chin. Phys. B 22(1) 018904
[12] Papadopoulos F, Kitsak M, Serrano M Á, Boguñá M and Krioukov D 2012 *Nature* **489** 537
[13] Yook S H Jeong H and Barabasi A L 2001 *Phys. Rev. Lett.* **86** 5835
[14] Eom Y H, Jeon C, Jeong H and Kahng B 2008 *Phys. Rev. E.* **77** 056105
[15] Barrat A, Barthelemy M and Vespignani A. 2004 *Phys. Rev. Lett.* **92** 228701
[16] Barrat A, Barthelemy M and Vespignani A. 2004 *Phys. Rev. E.* **70** 066149 arXiv:cond-mat/0406238
[17] Hanaki N, Peterhansl A, Dodds P S and Watts D J 2007 *Management Science.* **53(7)** 1036
[18] Ergun G, Rodgers G J 2002 *Physica A* **303** 261
[19] Zheng D F, Trimper S, Zheng B and Hui P M 2003 *Phys. Rev. E.* **67** 040102
[20] Dorogovtsev S N and Mendes J F F 2004 arXiv: cond-mat/0408343v2.